\documentclass{ieeeaccess}
\usepackage{cite}
\usepackage{amsmath,amssymb,amsfonts}
\usepackage{algorithmic}
\usepackage{graphicx}
\usepackage{textcomp}
\usepackage{multirow}
\usepackage{color,soul}
\usepackage{enumerate,mdwlist}
\def\BibTeX{{\rm B\kern-.05em{\sc i\kern-.025em b}\kern-.08em
    T\kern-.1667em\lower.7ex\hbox{E}\kern-.125emX}}
\begin{document}
\history{}
\doi{}

\title{Sequence Alignment Algorithm for Statistical Similarity Assessment}
\author{\uppercase{Jakub Nikonowicz} \authorrefmark{1},
\uppercase{Łukasz Matuszewski} \authorrefmark{1},\uppercase{Pawel Kubczak} \authorrefmark{1}}
\address[1]{The Faculty of Computing and Telecommunications, Poznan University of Technology, 60-965 Poznan, Poland}

\tfootnote{This work was supported by the Polish Ministry of Science and Higher Education under the status activity task 0314/SBAD/0205 in 2020.}

\markboth
{J. Nikonowicz \headeretal: Sequence Alignment Algorithm for Statistical Similarity Assessment}
{J. Nikonowicz \headeretal: Sequence Alignment Algorithm for Statistical Similarity Assessment}

\corresp{Corresponding author: Jakub Nikonowicz (e-mail: jakub.nikonowicz@put.poznan.pl).}

\begin{abstract}
This paper presents a new approach to statistical similarity assessment based on sequence alignment. The algorithm performs mutual matching of two random sequences by successively searching for common elements and by applying sequence breaks to matchless elements in the function of exponential cost. As a result, sequences varying significantly generate a high-cost alignment, while for low-cost sequences the introduced interruptions allow inferring the nature of sequences dependence. The most important advantage of the algorithm is an easy interpretation of the obtained results based on two parameters: stretch ratio and stretch cost. The operation of the method has been simulation tested and verified with the use of real data obtained from hardware random number generators. The proposed solution ensures simple implementation enabling the integration of hardware solutions, and operation based on only two sequences of any length predisposes the method to online testing.
\end{abstract}


\titlepgskip=-15pt

\maketitle

\section{Introduction}
\label{sec:introduction}
\PARstart{S}{equence} analysis is a broad and important field of science and technology, where digital signal processing in information engineering is at the forefront. In this area, the sub-field of comparing random sequences, although developed over the past years \cite{Blum1983}, is recently gaining importance particularly quickly, creating a separate domain in cryptography \cite{Wisniewska2018}. Therefore, a special case of sequence analysis is a comparison of two sequences to answer the question, how are they similar to each other and what is the peculiarity of their mutual relationship. Due to the increasing demand for comparative analysis, the discipline has developed tremendously in recent years \cite{Yutao2014,Popereshniak2019,Stankovic2019,Li2020}. There are many different methods of assessing sequence alignment and similarity, the most intuitive of which still seems to be the correlation and mutual information between sequences \cite{Horan2005,Jain2016}. Both methods are widely used in telecommunications and cryptography to assess, e.g., randomness of a signal \cite{Horan2005,Compagner1991,Gray1977}. However, they are of limited use when simultaneous analysis of the temporal and statistical properties of data sequences is required. As cryptographic systems play an important part in various applications, e.g., ensuring secure data transmission, it is imperative that the provided sequences of random numbers are of appropriate quality, i.e. expected to be possibly close to true randomness. Therefore, sequences are subject to statistical testing, e.g. defined in NIST SP 800-22 \cite{Rukhin2010} or AIS-31 \cite{Schindler2003} standards. However, statistical tests do not distinguish between truly random and high-quality pseudo-random sequences. As such, they cannot be used, e.g., to assess within a reasonable time whether the generator is under attack. A convenient solution for examining the determinism is the mechanism proposed in \cite{Dichtl2007}, based on repeatedly starting the generator with the same initial conditions. The described testing technique is called the restart mechanism and can help to check if the generator produces sequences as a result of deterministic or non-deterministic phenomena. On the collected data set, the chi-square test \cite{Jessa2011}, standard deviation \cite{Xu2016}, or entropy estimation is applied \cite{Turan2018}. However, the restart mechanism and following statistical tests require collecting many data sequences, and thus are difficult to apply in online sequence comparison and generator testing. Therefore, obtaining a low-data demanding yet efficient method that processes the currently obtained string against a single reference and returns easy-to-evaluate information about the relationship of sequences remains an open issue.

The solution to the considered problem may be sought in areas facing similar challenges. The problem of assessing how two sequences are similar to each other is also known in the analysis of natural language processing \cite{Searls2001,Pestian2012}. The quantification of the similarity between texts is not unique and unambiguous, and largely depends on the relative importance attached to individual particles, letters, words, phonemes, and grammar, and even on the general context of its occurrence. More detailed descriptions of the methods can be found in the work \cite{Vinga2003}. A related issue is also raised in the science of molecular biology, where sequence comparison is required in comparing primary biological sequence information \cite{Fuchs2002}, such as protein, amino-acid, DNA, RNA, etc. Methods used in biological research are mainly based on word frequency, distances defined in Cartesian space by frequency vectors, and information contained in the frequency distribution \cite{Gusfield1997}. Statistical algorithms involve, i.a., implicit Markov models, Bayesian methods for hypothesis testing, Kolmogorov's complexity and chaos theory \cite{Li2001}. What is more interesting in the context of considered problem is, however, a narrow group of sequence alignment methods. These methods are used in arranging molecular sequences to identify regions of similarity which may be a consequence of functional, structural, or evolutionary relationships between the sequences. The use of alignment comparison appears in numerous bioinformatics applications related to searching for a template in a database, where similarity is used to infer congruent structure or function \cite{Pearson2000}.

\begin{figure}[!h]
    \centering\includegraphics[scale=0.35]{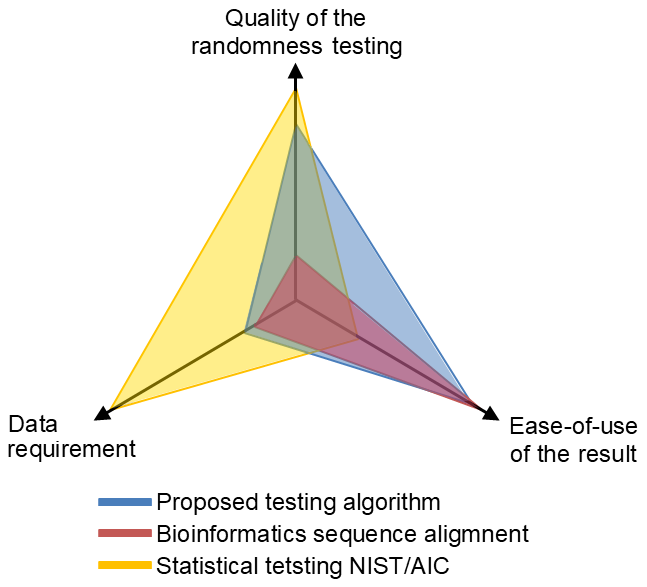}
    \caption{Characteristics of the proposed algorithm in comparison with available solutions. }
    \label{fig:0}
\end{figure}

Motivated by the desired functionality of random sequence testing and inspired by the above-mentioned algorithms, we propose an innovative algorithm for sequence comparative analysis. It combines the features of alignment, alignment-free, and information theory sequence comparison techniques, and is designed to compare random sequences produced by the random number generator - one of the most important elements required in cryptographic systems, and based on a small set of input data return a reliable, easy-to-evaluate result (Fig.~\ref{fig:0}).
The algorithm proposed in this paper requires only two sequences of any length and tries to match one to the other by applying sequence breaks to matchless elements in the function of exponential cost. As a result, sequences derived from a non-deterministic phenomenon generate a high-cost alignment, while for a low-cost sequences the introduced interruptions allow inferring the nature of the dependence between the sequences. After applying the algorithm, the results are obtained in the form of two metrics: stretch ratio and stretch cost, which makes the proposed solution easy to use and interpret. Moreover, the algorithm returns the stretched sequences, providing readily available material for further match and mismatch statistical analysis.

The rest of the article is organized as follows. Section~\ref{sec:2} describes the proposed alignment algorithm. Section~\ref{sec:3} explains the experiment methodology and shows numerical results. Finally, Section~\ref{sec:4} gives the concluding remarks.

\section{Algorithm description}
\label{sec:2}
\begin{figure}[!h]
    \centering\includegraphics[scale=0.55]{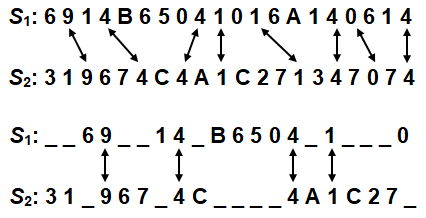}
    \caption{An example of two random sequences with common elements (top) adjusted by the algorithm (bottom).}
    \label{fig:1}
\end{figure}

\begin{figure}[!h]
    \centering\includegraphics[scale=0.55]{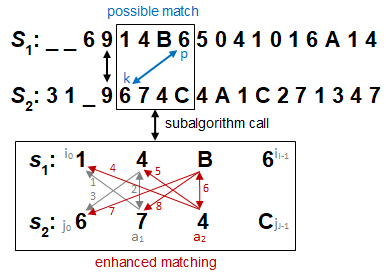}
    \caption{An example of the main algorithm operation with a call to enhanced matching subalgorithm.}
    \label{fig:2}
\end{figure}

\begin{figure}[!h]
    \centering\includegraphics[scale=0.63]{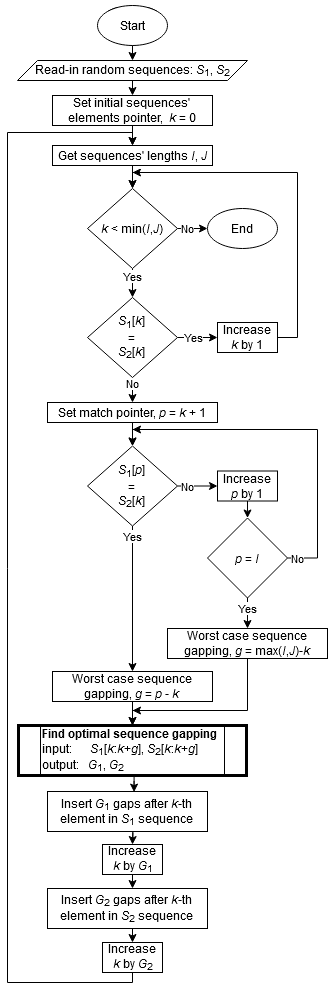}
    \caption{Block diagram of the main algorithm for two-way sequence alignment.}
    \label{fig:3}
\end{figure}

\begin{figure}[!h]
    \centering\includegraphics[scale=0.59]{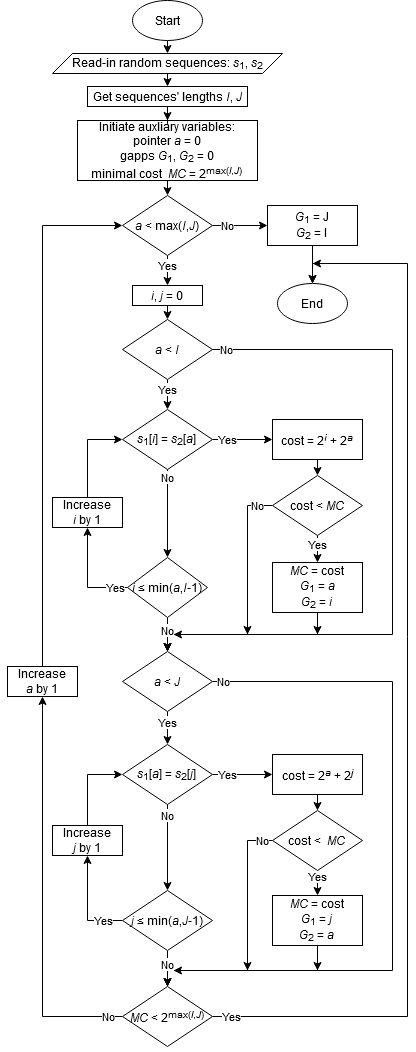}
    \caption{Sub-algorithm for finding optimal sequence gapping in the function of exponential cost.}
    \label{fig:4}
\end{figure}

The algorithm proposed in the paper performs mutual matching of two random sequences by successive searching for common elements in both sequences and inserting gaps for matchless ones (Fig.~\ref{fig:1}). An important feature of the algorithm is the careful minimization of the introduced gaps. The algorithm starts with the collection of two random sequences $S_{1}$ and $S_{2}$. Then it successively compares the elements of both sequences. In the case of a match, i.e., $S_{1}[k]=S_{2}[k]$, it increments index k and simply goes to the next comparison. In the case of a mismatch, i.e., $S_{1}[k] \neq S_{2}[k]$, it introduces an auxiliary index $p=k$. Then it begins the search through the sequence $S_{1}$ by increasing index $p$, so as to find an element matching the current element of $S_{2}$, i.e., $S_{1}[p]=S_{2}[k]$, where $p>k$. Such an initial match generated in a straightforward way would require the recognition that $p-k$ elements from $S_{1}$ do not have their counterparts in $S_{2}$, so the latter should be filled with $g=p-k$ gaps. The above represents the worst possible case where gaps are inserted in series only into sequence $S_{2}$. Therefore, the algorithm checks whether any of the elements preceding $S_{1}[p]$, i.e., those falling within the range k to p, match any element following $S_{2}[k]$ in the same range of indexes. The further operation of the algorithm assumes transferring both fragments of sequences, that is $s_{1}=S_{1}[k \ldots p]$ and $s_{2}=S_{2}[k \ldots p]$, to a sub-algorithm searching for the optimal gapping in both subsequences (Fig.~\ref{fig:2}). Detailed block diagrams of the alignment  algorithm and the sub-algorithm for optimal sequences gapping are presented in Figures~\ref{fig:3} and~\ref{fig:4}.

At this point it is worth noting that the search of the pair $S_{1}[p] = S_{2}[k]$ is primarily aimed at limiting the length of the sequences passed to the sub-algorithm. There is no limit to increasing index $p$, and in a critical case, e.g., searching for matchless element $S_{2}[k]$, the rest of both sequences will be passed to the optimizing sub-function. In the paper, the same set of values of both sequences is assumed, thus the presence of unique elements is minimized. However, the range of the search can be arbitrarily limited.

Searching for a better match of elements than the first in $s_{2}$ with the last in $s_{1}$ requires creating an appropriate cost function. The defined function should favour interrupting both sequences evenly, rather than introducing long series of gaps in just one of them. Therefore, the proposed sequence gapping assumes an exponential increase in the cost for series of gaps, i.e., for $G_{i}$ consecutive gaps the cost equals $2^{(G_{i}-1)}$. The total cost of gap insertion into both sequences is therefore equal to $2^{(G_{1}-1)} + 2^{(G_{2}-1)}$.

The sub-algorithm starts by reading both subsequences $s_{1}$, $s_{2}$ and determining their current lengths $I$ and $J$ respectively. Note that in the case of continually extending sequences $S_{1}$ and $S_{2}$ their final lengths do not have to be equal. Thus the lengths of last sub-sequences may differ.

Initially, the algorithm determines the maximum cost of matching, i.e., for $s_{1}[I-1]=s_{2}[0]$. Then, for a rolling index a pointing the current reference element, i.e., $s_{2}[a]$ or $s_{1}[a]$, it searches for a match from the opposing sequence. The found element, with the lowest possible index $i$ or $j$ preceding $a$, that is $s_{1}[i]$ or $s_{2}[j]$, respectively, is equivalent to the minimization of the gapping cost.

An example of operation of the sub-algorithm is shown in Figure (\ref{fig:2}). An initial match $S_{1}[p]=S_{2}[k]$ would require the insertion of three gaps into $S_{2}$ between elements 9 and 6 with the cost of $2^{2}$. The algorithm performs the search through subsequences $s_{1}$ and $s_{2}$ for a lower-cost match by a rolling comparison, i.e., $s_{1}[0]$ with $s_{2}[1]$, $s_{2}[1]$ with $s_{1}[1]$ and $s_{2}[0]$ with $s_{1}[1]$, and further $s_{1}[0]$ with $s_{2}[2]$, $s_{1}[1]$ with $s_{2}[2]$ etc. The approach enables it to find an optimal match, i.e., $s_{1}[1]=s_{2}[2]$, requiring the insertion of only one gap into $s_{1}$ and two gaps into $s_{2}$, thus generating the cost of $2^{0} + 2^{1}$. The determined numbers of the gaps, marked as $G_{1}$ and $G_{2}$, are returned to the main algorithm.

At this point, it is worth noting that the sub-algorithm is complemented by conditions ensuring correct searching in the case of subsequences of different lengths. Moreover, in order to avoid the occurrence of large numbers, while determining the exponential cost function, for long series of gaps the operation may be performed on the exponents only. In this case, the algorithm will test the condition

\begin{equation}
    i+a<2\left(c-1\right) \label{eq:1}
\end{equation}

\noindent and alternatively

\begin{equation}
    a+j<2\left(c-1\right) \label{eq:2}
\end{equation}

\noindent where $c$ is the exponent of the current minimum cost, i.e., $MC = 2^{c}$.

The main algorithm receives information on $G_{1}$, adds an adequate number of gaps after element $S_{1}[k]$ and updates the index to the position $k=k+G_{1}$. Then it receives the information on $G_{2}$ and inserts an adequate number of gaps after $S_{2}[k]$. This way, both sequences are stretched so that gaps in $S_{1}$ are inserted at the mismatch positions of $S_{2}$, and gaps in $S_{2}$ are introduced for mismatched elements in already stretched $S_{1}$.

The final goal of the algorithm is to obtain two pairs of metrics $S_{i}(a_{i},b_{i})$, describing each of the tested sequences:

\begin{itemize}
    \item $a_{i}$ is a stretch ratio determined as the ratio of the length of the gapped sequence $L_{i}^{'}$ to its initial length $L_{i}$. Importantly, in both lengths, only the number of elements to be matched is taken into account (Figure 5). As a result of the algorithm, the sequences increase in lengths dynamically. Consequently, in one of the sequences, a "tail" of length $L_{t}$ may be created. $L_{t}$ is the number of elements for which there were no more elements to match in the opposite sequence. Considering the above, the stretch ratio is redefined as
    
    \begin{equation}
        a_{i}=\left(L_{i}^{'}-L_{t}\right)/\left(L_{i}-L_{t}\right) \label{eq:3}
    \end{equation}
    
    \item $b_{i}$ is the cost of stretching the sequence, calculated as
    
    \begin{equation}
        b_{i}=\sum_{j}2^{G_{i}^{j}-1}/\left(L_{i}-L_{t}\right) \label{eq:4}
    \end{equation}
    
    where $j$ indexes successive series of gaps, and $G_{i}^{j}$ is equal to the number of gaps in the $j$-th series.
\end{itemize}

\begin{figure}[!h]
    \centering\includegraphics[scale=0.55]{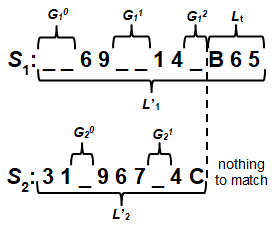}
    \caption{Sequences matched by the algorithm.}
    \label{fig:5}
\end{figure}

Figure~\ref{fig:5} and equations (\ref{eq:5}) to (\ref{eq:8}) show the process of computing the metrics for short sample sequences $S_{1}=[6,9,1,4,B,6,5]$ of the initial length $L_{1}$ and $S_{2}=[3,1,9,6,7,4,C]$ of the initial length $L_{2}$.

The metric for the first sequence $S_{1}(a_{1},b_{1})$ in the above example would be calculated as

\begin{equation}
    a_{1}=\left(L_{1}^{'}-L_{t}\right)/\left(L_{1}-L_{t}\right) \label{eq:5}
\end{equation}

\begin{equation}
    b_{1}=\frac{2^{G_{1}^{0}-1}+2^{G_{1}^{1}-1}+2^{G_{1}^{2}-1}}{L_{1}-L_{t}} \label{eq:6}
\end{equation}

\noindent whereas for the second sequence $S_{2}(a_{2},b_{2})$

\begin{equation}
    a_{2}=L_{2}^{'}/L_{2} \label{eq:7}
\end{equation}

\begin{equation}
    b_{1}=\frac{2^{G_{2}^{0}-1}+2^{G_{2}^{1}-1}}{L_{2}} \label{eq:8}
\end{equation}

Both metrics have their strict limits. The maximum extension of one sequence may be the number of elements of the other. Therefore, $S_{1}(a_{1},b_{1})$ metrics should be in the range

\begin{equation}
    a_{1} \in \langle 1,\frac{L_{1}+L_{2}-L_{t}}{L_{2}} \rangle \label{eq:9}
\end{equation}

\begin{equation}
    b_{1} \in \langle 0,\frac{2^{L_{2}-1}}{L_{1}-L_{t}} \rangle \label{eq:10}
\end{equation}

\noindent An analogous relationship with the length of $S_1$ takes place in the case of $S_{2}(a_{2},b_{2})$.

Considering the operation of the alignment algorithm, i.e., preferring interrupting both sequences evenly, the obtained $S_{i}(a_{i},b_{i})$ metrics are expected to be similar. Therefore, it is convenient to interpret a collective metric $S(a,b)$, where 

\begin{equation}
    a=\frac{L_{1}^{'}+L_{2}^{'}-L_{t}}{L_{1}+L_{2}-L_{t}} \label{eq:11}
\end{equation}

\begin{equation}
    b=\frac{\sum_{i}\sum_{j}2^{G_{i}^{j}-1}}{L_{1}+L_{2}-L_{t}} \label{eq:12}
\end{equation}

\begin{figure}[!h]
    \centering\includegraphics[scale=0.5]{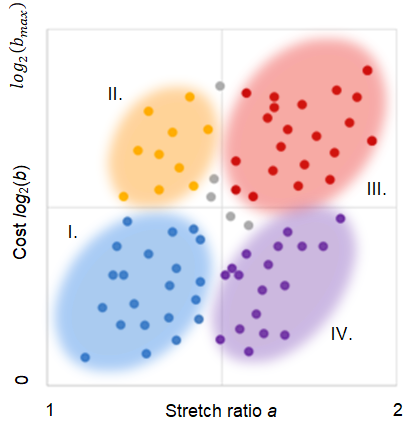}
    \caption{The meaning groups of the $S(a,b)$ metric in the plane of stretch and cost.}
    \label{fig:6}
\end{figure}

\noindent In the above case, the stretch ratio \begin{math} a \in \langle 1,2 \rangle \end{math} is a measure informing about the mutual similarity of both random sequences, whereas the cost of stretch \begin{math} b \in \langle 0,\left(2^{L_{1}-1}+2^{L_{2}-1}\right)/\left(L_{1}+L{2}\right) \rangle \end{math} informs about the average cost of processing each element, where a match costs 0. The described ranges reach given maximum values when all elements of the sequences are aligned and no tail is formed, i.e., $L_{t}=0$. We can interpret the collective metric $S(a,b)$ in an $a,b$ plane, as shown in Figure~\ref{fig:6}.

The metric can be classified into one of four general informative groups:
\begin{enumerate}[I]
    \item low stretch ratio indicates high similarity of sequences. Low cost is indicative of scattered and short interruptions, therefore rare mismatches. 
    \item low stretch ratio again means high similarity, while high cost informs about the insertion of gaps in series, i.e. cyclical convergences and divergences of both sequences.
    \item high stretch ratio and high cost are the result of introducing long series of gaps interleaved with sparse matches, thus informing about the negligible similarity of both sequences.
    \item high stretch ratio while maintaining low cost indicates frequent insertion of single gaps, i.e., scattered short matches and mismatches. The result is a premise for inference about the statistical similarity of the sequences.
\end{enumerate}

\section{Experiment}
\label{sec:3}
To provide proof of the concept, test data were prepared. Each test sequence has the length of 5000 elements, which take values from 0 to 15, and it corresponds to a four-bit representation. Section~\ref{prng} presents the comparisons of pseudorandom sequences and section~\ref{trng} shows the comparisons of sequences from hardware random number generators based on a Fibonacci Ring Oscillator (FIRO) \cite{b10}.

\subsubsection{PSEUDORANDOM SEQUENCES}
\label{prng}
\begin{figure}[!h]
    \centering\includegraphics[scale=0.5]{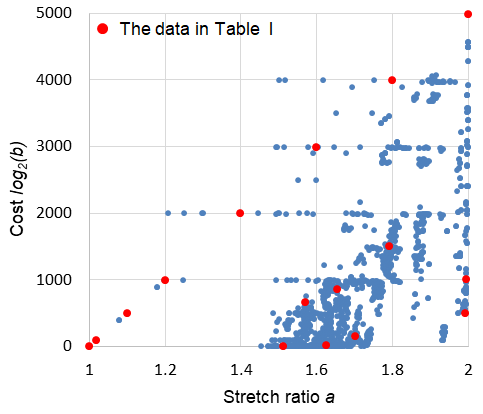}
    \caption{The pseudorandom sequences comparison results in the stretch-cost plane.}
    \label{fig:7}
\end{figure}

In the first step of the test, pseudorandom number sequences were generated with specific distributions. Four distributions were chosen: Gaussian, Uniform, Rayleigh, and Poisson. Sequences with the Gaussian distribution were generated with the mean value equal to two, four, eight, and fourteen. In sequences with the Uniform distribution, values from zero to fifteen occur with the same probability. For the Poisson distribution, the lambda parameter equal to two was chosen and for the Rayleigh distribution, the sigma parameter equal to two was set up. Compliance with the prepared data against intended results was checked by plotting and verifying histograms of the sequences. Additionally, to achieve certain statistic and time properties, some manual modifications of the sequences were made, i.e., in a sequence with the Gaussian distribution with the mean value of two and sigma of one, non-matching series of elements were inserted, and the extended length was truncated back to 5000. The used non-matching value was seven, as it did not occur in the original file. Insertions of lengths 100, 500, 1000, 2000, 3000, and 4000 were made. In total, 40 different files were prepared to check the properties of the metrics. Then the cost metric and the stretch ratio were calculated for all possible combinations of the files. Figure~\ref{fig:7} presents all 1600 results in the stretch-cost plane. Data presented in Table~\ref{tab:1} show how the inserted non-matching elements influence the cost and the stretch ratio. The dots highlighted in red show the data presented in Table~\ref{tab:1}.

\begin{table}[!t]
\centering
    \caption{Experiment results of pseudorandom sequences comparison.}
    \label{tab:1}
    \begin{tabular}{|cc|cc|}
    \hline
       \multicolumn{2}{|c|}{\begin{tabular}{@{}c@{}}Sequence in comparison \\ with Gauss $N$(2,1)\end{tabular}} & \multicolumn{2}{|c|}{Results}\\
    \hline
        Distribution &Comments &Stretch ratio $a$ &Cost $log_{2}(b)$\\
    \hline
        Gauss $N$(2,1) &0 INS &1 &0\\
        Gauss $N$(2,1) &100 INS &1.02 &86.71\\
        Gauss $N$(2,1) &500 INS &1.10 &486.71\\
        Gauss $N$(2,1) &1000 INS &1.20 &986.71\\
        Gauss $N$(2,1) &2000 INS &1.40 &1986.71\\
        Gauss $N$(2,1) &3000 INS &1.60 &2986.71\\
        Gauss $N$(2,1) &4000 INS &1.80 &3986.71\\
        Rayleigh $R$(2)  &- &1.51 &3\\
        Poisson $P$(2) &- &1.57 &654.71\\
        Uniform $U$(0,15) &- &1.62 &12.73\\
        Rayleigh $R$(3) &- &1.65 &853.71\\
        Gauss $N$(4,1) &- &1.70 &144.71\\
        Uniform $U$(0,15) &- &1.79 &1504.71\\
        Gauss $N$(8,1) &- &1.99 &491.71\\
        Gauss $N$(8,1) &- &1.99 &1009.71\\
        Gauss $N$(14,1) &- &2 &4985.76\\
    \hline
    \multicolumn{4}{c}{$\Delta F^{i}$ses; INS - in the sequence of certain distribution}\\
    \multicolumn{4}{c}{some non-matching insertions were made.}
    \end{tabular}
\end{table}

The cost corresponds to the size of the mismatched series indicated by continuous gaps, while increasing the gap size increases the stretch ratio, which is inversely proportional to the percentage of matches.  For example, when all values match, the stretch ratio is equal to one, and when there are no matches, it is equal to two.

A comparison of different sequences with Gaussian distributions shows that when the mean value difference is small, e.g., for a mean of two and mean of four, the stretch ratio remains low, but when the difference grows, and thus fewer values of both distributions overlap, then the stretch ratio also increases, reaching a maximum for the most distant pair, i.e., with the mean of two and the mean of fourteen.
The effect is visible for different distributions as well, e.g., the sequence with Gaussian distribution and the  mean of two is similar to data with Rayleigh distribution and sigma of two, therefore the stretch ratio for this pair is relatively low.

Data points presented in Figure~\ref{fig:7} confirm the interpretation of the groups presented in Figure~\ref{fig:6}. The stretch ratio shows the coincidence of both sequences, while the cost describes the matching pattern. Low cost indicates the insertion of short gaps series, and high cost informs about long breaks. Figure~\ref{fig:7} shows that for a stretch ratio close to maximum, the cost varies from 500 to 5000.

On the other hand, for the cost of 2000, the stretch ratio varies from 1.2 to 2. Thus, the obtained results fit into the predicted state-space in the a, b plane.

\subsubsection{TRUE RANDOM SEQUENCES}
\label{trng}
\begin{figure}[!h]
    \centering\includegraphics[scale=0.5]{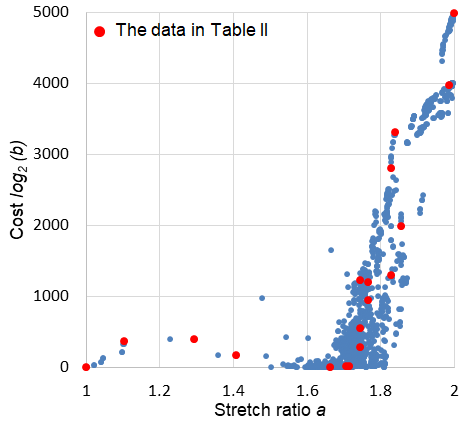}
    \caption{The FIRO random generator comparison results in the stretch-cost plane.}
    \label{fig:8}
\end{figure}

The second part of the experiment was focused on testing random sequences obtained from hardware random number generators. A major part of the sequences comes from a hardware generator implemented in an FPGA – Spartan 6 XC6LX16. The data set is supplemented with pseudo-random sequences from the Xilinx ISE Design Suite circuit simulator. Again, 40 different sequences were prepared, and all possible combinations of the pairs were examined. All results are presented in Figure~\ref{fig:8} and the data in Table~\ref{tab:2} are highlighted in red.

\begin{table}[!t]
\centering
    \caption{Experiment results of true random sequences comparison.}
    \label{tab:2}
    \begin{tabular}{|cc|cc|cc|}
    \hline
       \multicolumn{2}{|c|}{Sequence 1} & \multicolumn{2}{|c|}{Sequence 2} & \multicolumn{2}{|c|}{Results}\\
    \hline
        Source &Id$^{i}$ &Source &Id$^{i}$ &Stretch ratio $a$ &Cost $log_{2}(b)$\\
    \hline
        FPGA &d3\_r1 &FPGA &d3\_r1 &1 &0\\
        FPGA &d3\_r1 &FPGA &d3\_r3 &1.29 &395\\
        FPGA &d3\_r1 &FPGA &d17\_r3 &1.70 &7.39\\
        FPGA &d3\_r1 &FPGA &d99\_r3 &1.74 &280\\
        FPGA &d3\_r1 &ISE &d4\_c3 &1.77 &940.85\\
        FPGA &d3\_r1 &ISE &d4\_c1 &1.83 &2793.85\\
        FPGA &d3\_r1 &FPGA &d25\_r1 &1.99 &3960\\
        FPGA &d25\_r1 &FPGA &d25\_r4 &1.10 &359\\
        FPGA &d5\_r4 &FPGA &d5\_r1 &1.40 &159\\
        ISE &d1\_c3 &FPGA &d49\_r4 &1.66 &1.58\\
        ISE &d1\_c3 &FPGA &d3\_r3 &1.72 &3.58\\
        ISE &d1\_c3 &FPGA &d17\_r1 &1.75 &1215.85\\
        ISE &d1\_c3 &ISE &d3\_c1 &1.75 &544.71\\
        ISE &d1\_c3 &FPGA &d99\_r4 &1.77 &1196.85\\
        ISE &d1\_c3 &FPGA &d5\_r1 &1.83 &1287.85\\
        ISE &d1\_c3 &FPGA &d25\_r1 &1.84 &3300.85\\
        ISE &d1\_c3 &ISE &d4\_c1 &1.86 &1980.71\\
        ISE &d1\_c3 &ISE &d2\_c3 &1.99 &4983.71\\
    \hline
    \multicolumn{6}{c}{$^{i}$ Meaning of phrases; according to files d-delay, r-restart, and c-clock.}
    \end{tabular}
\end{table}

Table~\ref{tab:2} shows that in the control case, i.e., self-comparison of a sequence, the match cost is 0 and the stretch ratio is 1. However, for different sequences obtained from the same generator, the stretch ratio remains low with the match cost lower than 500. Careful analysis of the results clearly indicates sequences similar to each other, i.e., those having a match cost lower than 100 and a stretch ratio close to 1.7. By analogy, totally different sequences reach the stretch ratio close to 2 and the cost higher than 3000. When to different pairs of sequences, the same stretch ratio is assigned, e.g., close to 1.75, it is possible to distinguish the nature of their mutual convergence using a second metric. In such a case, the match cost varies from 0 to 1500. Using two metrics simultaneously provides a better overview of the similar type of both sequences.

The experimental results described above cover different types of dependencies, and once more confirm the expected performance of the proposed algorithm.

The results provided in Tables~\ref{tab:1} and~\ref{tab:2} demonstrate the wide application of the algorithm. The proposed method can be an effective tool for direct sequence comparison, and thus randomness assessment. Moreover, it can be used as a selector of structures and constructions which provide the highest match cost and stretch ratio, to ensure the high quality of generated random sequences. The metric can also be used as a general-purpose verification tool to compare the simulation results with the gathered real data.

\section{Conclusion}
\label{sec:4}
The algorithm described in the paper provides the results of a comparative analysis in the form of the $S(a,b)$ metric, which primarily ensures easy interpretation. The returned metric indicates the degree of mutual similarity of the sequences and allows one to infer the nature of their dependence. The expected behavior of the algorithm was confirmed both by simulation, using strings with known properties, and experimentally with the use of hardware random number generators with a known structure.

An important feature of the algorithm is its simple implementation, based on iterations, increments, and comparisons. This provides easy hardware realization as a build-in block in FPGAs and ASICs, whereas the use of a single reference enables the currently obtained sequence to be examined even in soft real-time systems. Therefore, the algorithm can be effectively used to follow the current operating state of the generator, and successfully used at the design stage as well, e.g., in the analysis of the similarity of sub-generators in more complex structures.

\begin{IEEEbiography}[{\includegraphics[width=1in,height=1.25in,clip,keepaspectratio]{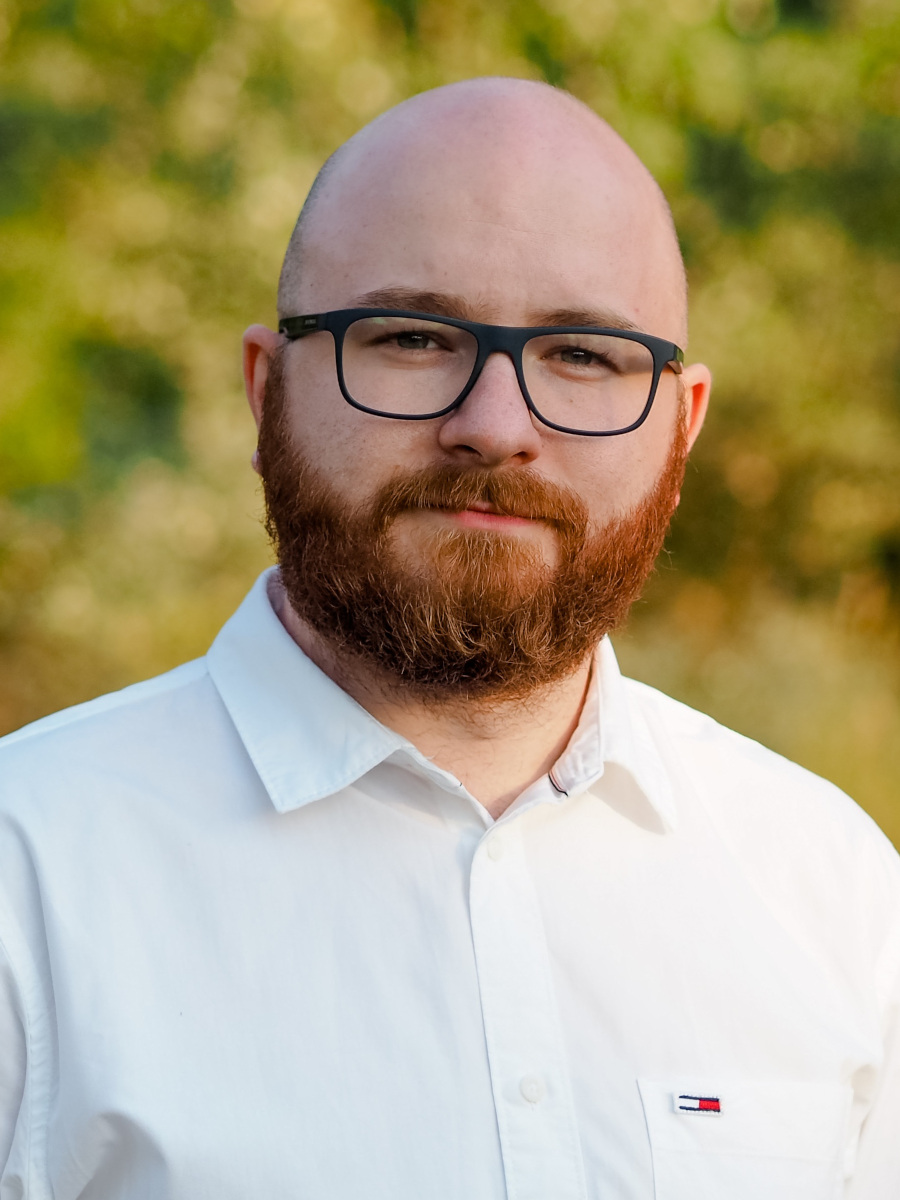}}]{Jakub Nikonowicz} was born in Poland in 1990. He received the M.Sc. degree in electronics and telecommunication and the Ph.D. degree (with honors) in telecommunication from the Poznan University of Technology, Poznan, Poland, in 2014 and 2019, respectively. He has authored or co-authored 10 scientific publications in refereed journals and proceedings of international conferences. His current research interests include statistical signal processing for blind signal detection and random number generation for reliable node authorization in wireless sensor networks.
\end{IEEEbiography}

\begin{IEEEbiography}[{\includegraphics[width=1in,height=1.25in,clip,keepaspectratio]{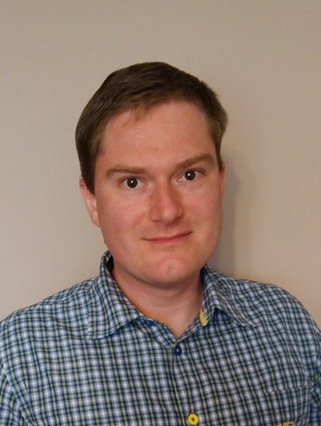}}]{Łukasz Matuszewski} graduated from the Faculty of Electronics and Telecommunications of the Poznań University of Technology in 2010. In 2011, he was employed at the Department of Telecommunications Systems and Optoelectronics at the Faculty of Electronics and Telecommunications of the Poznań University of Technology. In 2019 he defended PhD thesis entitled "The use of reprogrammable circuits to generate random sequences". From 2020, he is an assistant professor at the Institute of Multimedia Telecommunications at the Faculty of Computer Science and Telecommunications at the Poznań University of Technology. He was a project manager for grants for young scientists, as well as the contractor in projects for a Polish telecommunications operator. He is the author and co-author of 25 scientific publications in peer-reviewed journals and materials from national and international conferences. His research interests include designing devices with the use of reprogrammable circuits, in particular cryptographic circuits, random number generators, and synchronization circuits.
\end{IEEEbiography}

\begin{IEEEbiography}[{\includegraphics[width=1in,height=1.25in,clip,keepaspectratio]{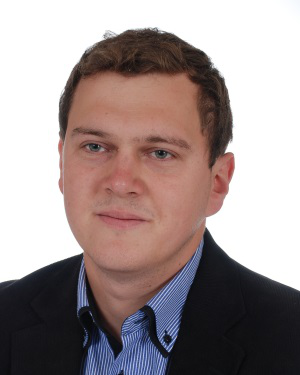}}]{Pawel Kubczak} graduated from the Faculty of Electronics and Telecommunications of the Poznań University of Technology in 2013. He continues his studies at the Faculty of Electronics and Telecommunications at the Poznań University of Technology. His research interests are related to digital random number generators, measuring the time period with picosecond accuracy and programmable digital circuits.
\end{IEEEbiography}

\EOD
\end{document}